\documentclass [aps,prl,twocolumn,groupeauthor,showpacs,superscriptaddress,longbibliography]{revtex4-1}

\usepackage{graphicx}% Include figure files
\usepackage{dcolumn}% Align table columns on decimal point
\usepackage{bm}% bold math$
\usepackage{helvet}
\usepackage{amssymb}
\usepackage{amsmath}
\usepackage{amsfonts}
\usepackage{hyperref}
\usepackage{color}

\begin{document}
  \title{Slow down of the electronic relaxation close to the Mott transition}
  \author{Sharareh Sayyad}

\affiliation{Max Planck Institute for the Structure and Dynamics of Matter, 22761 Hamburg, Germany}
\affiliation{University of Hamburg-CFEL, 22761 Hamburg, Germany}

\author{Martin Eckstein}
\affiliation{Max Planck Institute for the Structure and Dynamics of Matter, 22761 Hamburg, Germany}
\affiliation{University of Hamburg-CFEL, 22761 Hamburg, Germany}
  
%\date{\today~\tt\jobname.tex}
\date{\today}
  
%ME please sort the references (in list of changes, enough to say we corrected a misorderng of the references)

\begin{abstract}
We investigate the time-dependent reformation of the quasiparticle peak in  a correlated metal near the Mott transition, after the system is quenched into a hot electron state and equilibrates with an environment which is colder than the Fermi-liquid crossover temperature.  Close to the transition, we identify a purely electronic bottleneck timescale, which depends on the spectral weight around the Fermi energy in the bad metallic phase in a non-linear way. This timescale  can be orders of magnitude larger than the bare electronic hopping time, so that a separation 
of
electronic and lattice timescales may break down. The results are obtained using nonequilibrium dynamical mean-field theory and a slave-rotor representation of the Anderson impurity model.
\end{abstract}

  
\maketitle
When the Mott metal-insulator transition \cite{Imada1998} is approached from the metallic side, a narrow quasi-particle band emerges at the Fermi energy, and spectral weight is transferred into the Hubbard bands. This behavior, which is observed in a large class of materials, is a paradigm manifestation of many-body correlations, and its theoretical description has been a major success of dynamical mean-field theory (DMFT) \cite{Georges1996, Kotliar2006}. By means of photo-excitation, metallic phases in Mott insulators can be induced on femtosecond timescales  \cite{Iwai2003, Perfetti2006, Wegkamp2014}, which provides an intriguing example for ultra-fast switching material properties. While it is well understood that an intense laser pulse can rapidly  promote electrons to effective temperatures of several $1000K$ and thus lead to a partial melting of the Mott gap \cite{Perfetti2006}, the equilibrium properties of such a high-temperature state would correspond to a bad metal rather than a Fermi liquid \cite{Merino2000, Deng2013}. It thus remains a fundamental question, with immediate importance for understanding the transport properties of photo-excited metallic states, how fast coherent quasi-particles can be formed as the excitation energy is passed from the electrons to the lattice.

Naively one may expect that the electrons in a metal thermalize to a quasi-equilibrium state almost instantly after the excitation, and quasiparticles are formed as soon as the effective temperature is low enough. The relevant timescale for this process would then be set by the electron-lattice relaxation. 
In this work we show that a rapid thermalization can fail even in the  metallic phase. While
thermalization can be understood within a quasiparticle picture (from a kinetic equation), the latter provides no clue about the timescale for the evolution of the density of states itself, as long as quasiparticles are not yet well-defined. Considerable progress in describing the dynamics of Mott insulators has been made using nonequilibrium DMFT \cite{Aoki2014}, but a study of the correlated metal close to the Mott transition has remained elusive. Although the quasiparticle peak within DMFT corresponds to the Kondo resonance in an effective impurity model \cite{Georges1996}, its formation in time can be expected to show an entirely different dynamical behavior that the classic problem of the  buildup of Kondo screening \cite{Nordlander1999, Anders2005, Roosen2008, Cohen2013, Cohen2014, Antipov2015}, because the spectral weight responsible for the Kondo screening is formed {\em self-consistently} in DMFT. Impurity solvers such as higher-order strong-coupling expansions \cite{Eckstein2010}, Monte Carlo \cite{Werner2009c}, or density-matrix renormalization group \cite{Wolf2014}  have not yet reached sufficiently long times in this parameter regime. 

In equilibrium, the slave-rotor approach developed by Florens and Georges \cite{Florens2002, Florens2004} provides an intuitive semi-analytical understanding of the Mott transition, by representing electrons in terms of a quantum rotor (charge) and a spinful fermion. In this paper we solve the coupled spinon and rotor equations out of equilibrium, and show that the two partial degrees of freedom  become almost decoupled during the evolution. As a consequence, bad metallic behavior prevails in a photo-excited state over times which can be orders of magnitude longer than the electron hopping, and therefore even become comparable to the electron-phonon relaxation time.

{\bf Model:} 
We study the particle-hole symmetric Hubbard model
\begin{equation}\label{eq:Hubbardmodel}
 H =
-J(t)\sum\limits_{\langle ij\rangle, \sigma}
(c^{\dagger}_{i\sigma}
c_{j\sigma}
+ h.c.)
+
U
\sum\limits_{i} 
(n_{i\uparrow}-\tfrac12)
(n_{i\downarrow}-\tfrac12)
,
\end{equation}
where $J(t)$ is the time-dependent hopping amplitude,
$U$ is the on-site Coulomb repulsion, $c_{i\sigma}$ and $c^{\dagger}_{i\sigma}$ are  electron annihilation and creation operators for spin $\sigma \in\{ \uparrow,\downarrow\}$ on site $i$, and $n_{i \sigma}=c_{i\sigma}^\dagger c_{i\sigma}$.
 To study the time-dependent formation of quasiparticles, we initially prepare the system in the atomic limit ($J=0$), and rapidly turn on the hopping to a value $J_0>0$. (In the following, $J_0$ and $\hbar/J_0$  set the energy and time unit, respectively, and the ramp-on profile is given by $J(t)= J_0( 1-\cos(\pi t/t_c) \big)/2$ for $ 0\le t \le t_c=2.5$.) 
 The model is solved within nonequilibrium DMFT \cite{Aoki2014} on a Bethe lattice, i.e., it is mapped onto an  Anderson impurity problem with self-consistently determined hybridization function $\Delta(t,t')=J(t)G_\text{loc}(t,t')J(t')$ \cite{Georges1996}, where  $G_{loc}(t,t')=-i\langle T_{\cal C}c(t) c^{\dagger}(t') \rangle$ is the local contour-ordered Green's function~\cite{footnote1}.

To solve the dynamics of this Anderson model, we employ the U(1) slave-rotor representation \cite{Florens2002}. 
 The impurity operators ($c_{\sigma}$, $c_{\sigma}^{\dagger}$) are substituted by $c_{\sigma}^{\dagger}=f_{\sigma}^{\dagger}e^{i\theta}$, where $f_{\sigma}^{\dagger}$ is a fermion and $\theta \in [0,2\pi)$ is a quantum rotor variable. A constraint $L=\sum_{\sigma} f_{\sigma}^{\dagger}f_{\sigma}-1$ on the angular momentum $L=i\partial_\theta$  of the rotor removes unphysical states from the Hilbert space. With this, the interaction Hamiltonian is determined only by the rotor, 
 $H_U=UL^2$, while $f_\sigma$ represents a charge-less fermion (spinon). Furthermore, the rotor is replaced by a bosonic field  $X=e^{i\theta}$ with the constraint  $|X(t)|^{2}=1$. 
The dynamics of the impurity model is then analyzed in terms of contour-ordered rotor and spinon Green's functions
\begin{subequations}
\label{GFX}
\begin{align}
\label{GX}
 G_{X}(t,t')&=-i\langle T_{\cal C}X(t) X^{*}(t') \rangle ,\\
 \label{Gf}
 G_{f}(t,t')&=-i\langle T_{\cal C}f_{\sigma}(t)f^{*}_{\sigma}(t') \rangle,
\end{align}
\end{subequations}
where $G_X$ has a direct relation to the local charge susceptibility \cite{Florens2002},
and subsequently the electron's Green's function is obtained by $G_{loc}(t,t')=iG_{f}(t,t')G_{X}(t,t')$.
The model can be solved exactly when the spin-degeneracy $N$ and the number of rotor flavors $M$ is increased from $N=2$ and $M=1$ to infinity, keeping the ratio $\mathcal{N}=N/M$ fixed \cite{Florens2002},  and this limit provides a qualitatively correct description of the metal-insulator transition. The resulting integral equations correspond to a reformulation of the Ref.~\cite{Florens2002} within the Keldysh framework and are given in the supplementary material \cite{Supplementary}.

\begin{figure}[tbp]
\centerline{ \includegraphics[width=0.99\columnwidth]{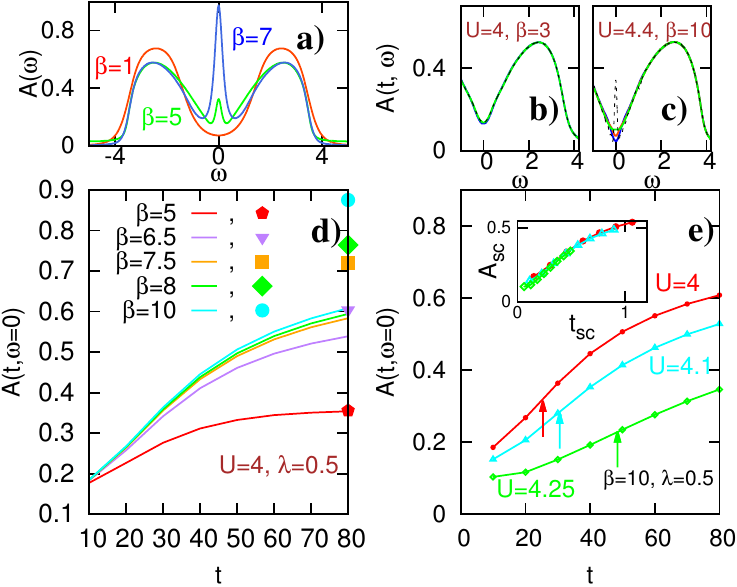}}
 \caption{ 
 a) Equilibrium density of state $A(\omega)$ for three different temperatures throughout the metal-insulator crossover at $U=4$. 
 b-c) $A(\omega)$ in equilibrium (dashed line) compared to the time-dependent spectral function $A(t, \omega)$  for $\lambda=0.5$ at times $t=20$ (blue), $t=50$ (red), and $t=80$ (green).
 d) The height of the quasiparticle peak $A(t,\omega=0)$ as a function of time for $U=4$, bath temperatures $\beta=5,6.5,7.5,8,10$ (from bottom to top) and $\lambda=0.5$.  Symbols on the right vertical axis correspond to the equilibrium value $A(\omega=0)$ at the same temperatures. e) The height of the quasiparticle peak  $A(t,\omega=0)$ as a function of time for $U=4, 4.1, 4.25$, bath temperature $\beta=10$, and $\lambda=0.5$. Arrows indicates the time 
$t_\text{max}$ (Fig.~\ref{fig:Gf_derGf}c).
The inset of e) plots the time-dependent spectral function $A_{\rm sc}(t_{\rm sc})=A(t, \omega=0)$  as a function of rescaled time $t_{\rm sc}= t/\tau^*$  with arbitrary rescaling of the vertical axis (see main text).
}
\label{fig:h_Kondo}
\end{figure}

{\bf Results:}  In Fig.~\ref{fig:h_Kondo}a, we plot the electronic density of states for three temperatures at $U/J_0=4$
in equilibrium (time-independent $J(t)=J_0)$. 
The metal insulator transition endpoint is at $U_{\rm c}\approx4.69$. Below a temperature $T^{*}\approx0.2$, a quasiparticle peak emerges at the Fermi energy, while for $T>T^*$ the system is in a bad metallic state with a pseudo-gap at the Fermi energy.
At intermediate values of $U$ and in an isolated system,
a $J(t)$-like
quench would lead to a highly excited electronic state which thermalizes within few $1/J_0$ to an effective temperature above the Fermi-liquid crossover $T^*$ \cite{Eckstein2011}, which is also confirmed by the slave-rotor calculations. Hence this excited state is a good representation of a hot-electron state reached after strong photo-excitation. In addition, the system is weakly coupled to a bosonic heat bath at low temperature $T=1/\beta$, to cool down the electrons and form the Fermi liquid when $U/J_0$ is in the metallic phase. We treat this dissipative bath by an additional electron self-energy $\Sigma_{bath}(t,t') = \lambda D(t,t')G(t,t')$, where $D$ is the noninteracting bosonic Green's function with frequency $\omega_{0}$, set to $\omega_{0}=1$, and $\lambda$ is the coupling constant \cite{Eckstein2013b}.
The coupling is small enough so that the effect of the bath on the electronic density of states is weak, and the bath provides only energy relaxation. 

To track the time-evolution of the system, we compute the time-dependent spectral functions $A(t, \omega) =-\frac{1}{\pi}\text{Im} \int_0^t ds \,G^\text{ret}(t,t-s)e^{i\omega s}$. For bath temperatures $T>T^*$, $A(t,\omega)$ is almost indistinguishable from the equilibrium spectrum $A(\omega)$ at temperature $T$ already at early times $t=20$ (Fig.~\ref{fig:h_Kondo}b). For lower temperature, however, only the Hubbard bands are rapidly retrieved, while the formation of the quasiparticle peak remains incomplete even for times larger than the inverse width of the peak 
(Fig.~\ref{fig:h_Kondo}c).
%ME no new parargraph
The slow evolution is also clear from the time-evolution of the spectral weight $A(t,\omega=0)$ 
(Fig.~\ref{fig:h_Kondo}d): For $T<T^{*}$, the equilibrium value $A(0)$ strongly increases with decreasing $T$, while the time-dependent value $A(t,0)$ becomes almost independent of the bath temperature, indicating that the dynamics is governed by a bottleneck of electronic nature.
 The closer $U$ is to the critical value $U_{\rm c}$, the less metallic is the transient state 
(Fig.~\ref{fig:h_Kondo}e).
(The presence of such a bottleneck makes it impossible to extrapolate the data $A(t,\omega=0)$ to the final equilibrium value from the early times around the bottleneck.)   We note that qualitatively the same behavior is found using the non-crossing impurity solver~\cite{Eckstein2010}, although the latter is not quantitatively accurate around the metal-insulator transition, and $U_c$ is reduced~\cite{Supplementary}.   In the following, we will use the slave-rotor language to identify a purely electronic crossover timescale which is related to the spectrum in a rather nontrivial way [c.f.~Eq.~\eqref{a2cosh}~below], and which captures the slow-down of the relaxation.

\begin{figure}[tbp]
\centerline{ \includegraphics[width=0.99\columnwidth]{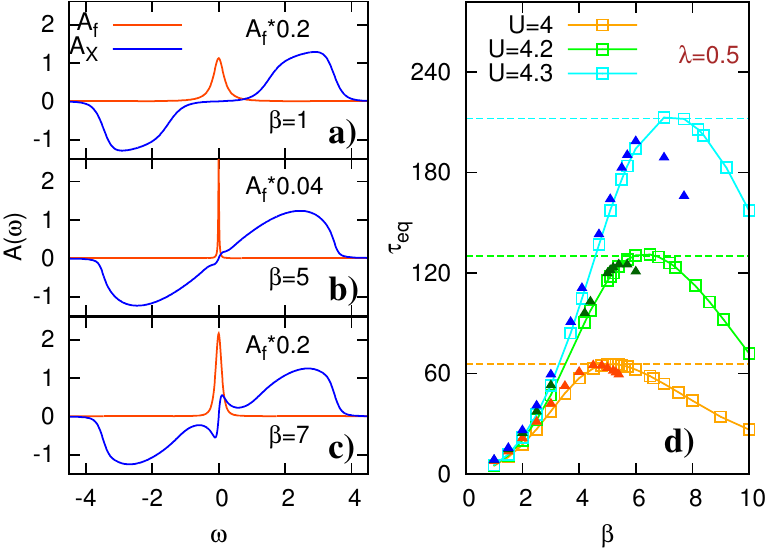}}
 \caption{
a)-c) Spinon ($f$) and rotor ($X$) spectral functions in equilibrium for $U=4$ and  temperatures in the bad metal regime ($\beta=1$, a), the crossover ($\beta=5$, b) and the metallic phase ($\beta=7$, c). d) The spinon lifetime  (inverse width of the peak)  as a function of $\beta$ for different values of $U$. Triangular points are calculated using the approximate expression Eq.\eqref{a2cosh}.  Dashed lines indicate  the maximum spinon lifetime $\tau_\text{max}$ during the relaxation process (see main text and Fig.~\ref{fig:Gf_derGf}).  }
\label{fig:Bandwidth}
\end{figure}

\begin{figure}[tbp]
\centerline{ \includegraphics[width=0.99\columnwidth]{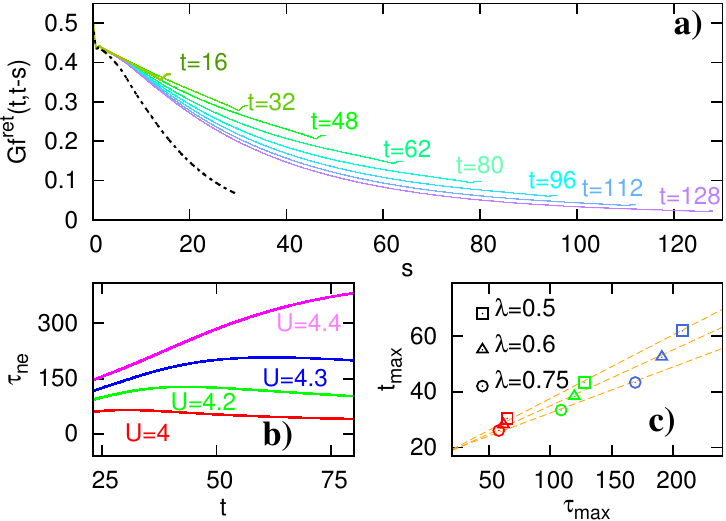}}
 \caption{
a) Retarded spinon Green's function $G_f^\text{ret}(t,t-s)$ as a function of relative time $s$ for 
various different times $t$ ($U=4$, $\beta=10$, $\lambda=0.5$), and in equilibrium (dashed line). 
The slope decreases for $t\lesssim 32$ and increases for $t \gtrsim 32$. 
b) Inverse of the slope
[Eq.~\eqref{tau}] 
as a function of $t$ for $\lambda=0.5$,  $\beta=10$, $s_0=16$ and various values $U$ below the metal-insulator transition;
%ME
$s_0=16$ is chosen large enough so that $G_f(t,t-s_0)$ reflects the low-energy part of the Green's function (the Lorentzian peak).
c) Crossover time $t_\text{max}$ plotted against $\tau_\text{max}$, where $(t_\text{max},\tau_\text{max})$ 
corresponds to the maximum of the curves $\tau_{\rm ne}(t)$ in panel b). }
\label{fig:Gf_derGf}
\end{figure}

 Despite the well-known equilibrium physics of the Hubbard model, the slave-rotor language exhibits a nontrivial spinon response in the crossover regime. Figure \ref{fig:Bandwidth}a-c shows the spectral functions $A_{X,f}(\omega)=-\frac{1}{\pi}\text{Im} \int ds \,G_{X,f}^\text{ret}(s)e^{i\omega s}$ in equilibrium. At high-temperature, the rotor has spectral weight around the Hubbard bands, and the spinon peak is broadened due to the interaction with the charge fluctuations (Fig.~\ref{fig:Bandwidth}a). Below the crossover (Fig.~\ref{fig:Bandwidth}c), the rotor develops low-energy spectral weight, which implies the formation of the quasi-particle peak \cite{Florens2002}. In the intermediate temperature regime, however, spinon and rotor become energetically 
weakly coupled,
and $A_f(\omega)$ develops into a narrow Lorentzian peak (Fig.~\ref{fig:Bandwidth}b). The width $\Gamma$ of the Lorentzian defines a timescale $\tau_\text{eq}(T)=1/\Gamma$, which has a clear maximum $\tau_*$ as a function of temperature in the metal-insulator crossover (Fig.~\ref{fig:Bandwidth}d)
~\cite{footnote-singularity}
. To characterize the time evolution, we plot $G_f^\text{ret}(t,t-s)$ as a function of time difference $s$ for various $t$ (Fig.~\ref{fig:Gf_derGf}). A narrow peak in $A_f(\omega)$ corresponds to a slow decay of $G_f$ as a function of $s$, so that we can define a nonequilibrium spinon lifetime by 
 \begin{equation}
 \label{tau}
  \tau_\text{ne}^{-1}(t)=-\partial_s G_f^{\rm ret}(t,t-s)/G_f^{\rm ret}(t,t-s)|_{s=s_0}
 \end{equation} 
 for some fixed time $s_0$.  (For a Lorentzian peak, $\tau$ is the inverse width).
 The time $\tau_\text{ne}$ first increases with $t$ and then decreases, tracking the evolution of $\tau_\text{eq}(T)$ a function of temperature (Fig.~\ref{fig:Gf_derGf}b). 
 We then find that,
 for a given coupling to the bath,
 the maximum of $\tau_{\rm ne}(t)$ as a function of time ($\tau_\text{max}$) coincides with the crossover scale $\tau_*$ (dashed lines in Fig.~\ref{fig:Bandwidth}d), and moreover, this value is reached at a time $t_\text{max}$ proportional to $\tau_*$ (Fig.~\ref{fig:Gf_derGf}c). The electronic spectral function at $t=t_{\rm max}$ is at the onset of quasiparticle formation~(arrows in Fig.~\ref{fig:h_Kondo}e).  
 This demonstrates that the long lifetime of the spinon provides a bottleneck time for the relaxation in the crossover regime.

In Fig.~\ref{fig:h_Kondo}e (and the inset) we show that the evolution of the quasiparticle peak height $A(t, \omega=0)$ in the crossover regime for various values of $U$ can be roughly collapsed on each other when the time axis is rescaled by this crossover scale $\tau_*$, i.e.,  to this extent $\tau_*$ determines also the slowdown of the electronic relaxation. Furthermore, 
although properties of $A_f(\omega)$ are not simply reflected in the equilibrium single-particle properties, one can approximately express the timescale $\tau_*$ in terms of the electronic degrees of freedom. The width of a sharp resonance in $A_f(\omega)$ is given by the imaginary part of the self-energy $\Sigma_f$, which depends on $G_X$ and $\Delta$. In the crossover regime, we can, to a first approximation, relate the rotor $G_X$ to the electronic Green's function by  $A(\omega) = \frac{1}{2} A_{X}''(\omega) \coth(\beta\omega/2)$, by setting $A_f(\omega)=\delta(\omega)$ in the convolution $G_{loc}(t,t')=iG_{f}(t,t')G_{X}(t,t')$
\cite{footnote02}.  Analytic continuation of $\Sigma_{f}(t,t')=i\Delta(t,t') G_{X}(t',t)$ then gives
\begin{equation}
\tau^{-1}_{\rm eq}(T)
=
-\Sigma_{f}''(\omega=0)
\approx
-\int d\omega
\,\frac{\Delta''(\omega)A(\omega)}{\cosh(\omega/2T)^2},
\label{a2cosh}
\end{equation}
and $\tau_*=\tau_{\rm eq}(T^*)$, which agrees well with the numerical result (Fig.~\ref{fig:Bandwidth}d). For the Bethe lattice $\Delta=J^2G_{loc}$, so that $\Delta''(\omega) = -\pi J^2 A(\omega)$. Equation \eqref{a2cosh} implies a rather nontrivial relation between the nonequilibrium relaxation and the electronic properties. At $T^*$, the hyperbolic cosine function restricts the integral to values close to the pseudo-gap, where $A(\omega)$ is small. Since the endpoint of the metal-insulator transition temperature in the Hubbard model is remarkably small compared to the bare energy scales, $\tau_*$ becomes much longer than the bare hopping close to the transition $U=U_c$.

\begin{figure}[tbp]
\centerline{ \includegraphics[width=0.99\columnwidth,clip=true]{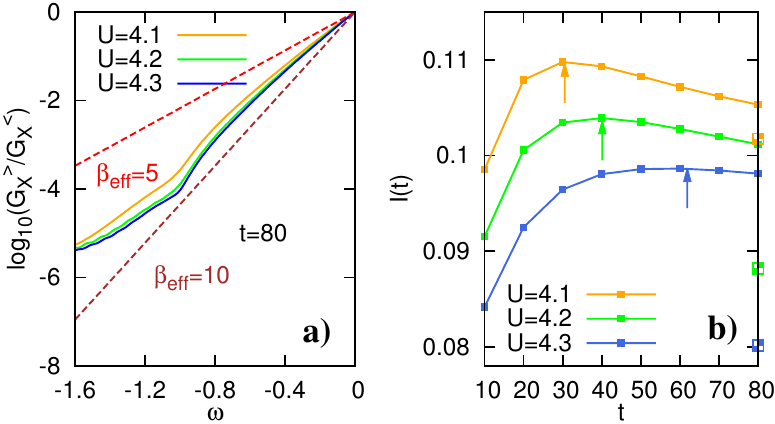}}
 \caption{
Test of the quasi-equilibrium relation $G_X^>(\omega,t)/G_X^<(\omega,t) = e^{\beta_\text{eff}\omega}$ at $t=80$ for $U=4,4.2,4.3$ ($\lambda=0.75$, $\beta=10$). 
b) Integrated spectral weight $I(t)=\int_{0}^{0.5} {\rm d\omega} A_X(t,\omega)$ of the rotor at low energy.  Square half-filled points at $t=80$ correspond to the equilibrium values. Arrows indicate the $t_{\text{max}}$ (Fig.~\ref{fig:Gf_derGf}c).
  }
\label{fig:Beta_fit_Xloc}
\end{figure}

To further analyze the relaxation, one can check whether rotor reaches a quasi-equilibrium state while the spinon is slowly evolving,
by testing whether the fluctuation-dissipation $G_X^>(\omega,t)/G_X^<(\omega,t) = e^{\beta_\text{eff}\omega}$ is satisfied. The latter  
would imply that an effective temperature $T_\text{eff}=1/\beta_\text{eff}$ can be assigned to charge fluctuations 
(using the time-dependent Fourier transforms $G^{>,<}(\omega,t) =\int ds \,G^{>,<}(t,t-s) e^{i\omega s}$). Figure \ref{fig:Beta_fit_Xloc}a  however shows that a single charge temperature cannot be defined on the timescale of the simulation. While the occupation of high-energy fluctuations (the Hubbard bands) is small, the  
 low energy part remains at an apparent higher temperature (the slope of lines for $\omega \gtrsim -0.8$ is slightly smaller than for $\omega \lesssim -0.8$), i.e., high-energy and low-energy charge fluctuations are not thermalized with each other.  The lower effective temperature for larger $U$ may be related to the lowering of the crossover temperature with increasing $U$.
Because of the coupling between the spinon and the rotor, the low-energy spectral weight of the rotor also reflects the non-monotonous evolution of the spinon (Fig.~\ref{fig:Beta_fit_Xloc}b): The increase of the spinon bandwidth for $t>t_\text{max}$ leads to the transfer of rotor spectral weight to higher energies, so that the integrated spectral weight $I(t)$ of the rotor in the low energy region $0< \omega  < 0.5$ has a maximum around $t=t_\text{max}$.

{\bf Conclusion and Discussion:} 
In conclusion, we have investigated how the electronic state close to the Mott transition in the Hubbard model relaxes from an excited hot-electron state towards the Fermi liquid. We found a bottleneck time of purely electronic nature, before which charge and electronic degrees of freedom remain in a non-thermal state and cannot be characterized by an effective temperature, and the formation of the quasiparticle band is incomplete. The electronic relaxation is related to the spinon lifetime $\tau_*$, and a simple estimate [Eq.~\eqref{a2cosh}]  in terms of the density of states around $\omega=0$ at the crossover temperature $T^*$ (the onset of quasiparticle formation) shows that this time can be much longer than the femtosecond hopping time,
and thus violates the paradigm of rapid thermalization in a metal.
The absence of quasiparticles implies long-lived bad metallic behavior, and should thus be observable also in optical experiments on materials like ${\rm LiV_{2}O_{4}}$, which are metals close to the paramagnetic Mott transition \cite{matsuno2000, Arita2007}.  We note that slow (or absent) formation of a quasiparticle band was also observed in simulations of a photo-doped Mott-insulator \cite{Eckstein2013b}, but in this case the origin of the behavior is less clear because the final low temperature state is insulating.

It is important to note that the $\tau_{*}$ characterizes the slow dynamics of the system around the crossover regime, but not necessarily the subsequent reshaping of the quasiparticle peak.
Times larger than $t_{\rm max}$ cannot be studied systematically due to the increase of the numerical cost with the simulated time.
The final formation of the quasiparticle peak might bring in another slow timescale related to the build-up of low energy spectral weight of the rotor.  
Furthermore, after quasiparticles are formed, slow dynamics can arise also from an ineffective coupling of heavy electrons to phonons \cite{Demsar2003}.
The observed dynamical behavior arises from the DMFT self-consistency and is thus a lattice effect. In contrast, the build-up of the Kondo peak after a quench in the Anderson model is limited only by energy-time uncertainty \cite{Nordlander1999} (i.e., the formation of the peak is complete after the inverse of its width), and this behavior is also reproduced by the slave-rotor method \cite{Supplementary}.
A natural question for future studies is thus whether a similar electronic bottleneck time may appear in multi-band Hubbard models or the Kondo lattice model for heavy fermions, where localized $f$ or $d$ orbitals interacting with delocalized electrons giving rise to the emergence of massive quasiparticles. In this context it is also interesting whether one can identify the small energy scale related to the spinon in numerically exact equilibrium calculations. 

The authors would like to acknowledge fruitful discussions with A. Rosch, D. Golez, and Ph. Werner.

\end{document}

\title{Supplementary Material for\\"{Slow down of the electronic relaxation close to the Mott transition}``}
  \author{Sharareh Sayyad}
	
\affiliation{Max Planck Institute for the Structure and Dynamics of Matter, 22761 Hamburg, Germany}
\affiliation{University of Hamburg-CFEL, 22761 Hamburg, Germany}

\author{Martin Eckstein}
\affiliation{Max Planck Institute for the Structure and Dynamics of Matter, 22761 Hamburg, Germany}
\affiliation{University of Hamburg-CFEL, 22761 Hamburg, Germany}

\date{\today}

\maketitle

\section{Anderson Impurity Model: slave-rotor decomposition}

In this section, we present some technical details of the nonequilibrium DMFT solution of the Hubbard model
\begin{equation}\label{eq:Hubbardmodel}
 H =
-J(t)\sum\limits_{\langle ij\rangle, \sigma}
(c^{\dagger}_{i\sigma}
c_{j\sigma}
+ h.c.)
+
U
\sum\limits_{i} 
(n_{i\uparrow}-\tfrac12)
(n_{i\downarrow}-\tfrac12)
,
\end{equation}
with the slave-rotor representation of the Anderson impurity model,
%ME
generalizing Ref.~\cite{Florens2002} to the Keldysh framework.
Here $J(t)$ is the time-dependent hopping amplitude,and $U$ is the Hubbard strength.

Within DMFT, the mapping of the lattice problem~Eq.~\eqref{eq:Hubbardmodel} onto the impurity model will lead us to the action~\cite{Eckstein2010b}
\begin{align}
 \mathcal{S}=& -i \int_{\cal C} {\rm dt dt'} c^{\dagger}(t) \Delta(t,t') c(t') +  \mathcal{S_{\rm loc}} \label{Eq:actionelectron}
\end{align}
where 
\begin{equation}
 \mathcal{S_{\rm loc}} = -i\int_{\cal C} {\rm dt} U \big( n_{\downarrow} -\frac{1}{2} \big)\big( n_{\uparrow} -\frac{1}{2} \big),
\end{equation}
 $\cal C$ stands for the Keldysh contour, $\Delta$ is the hybridization function of electrons $\Delta(t,t')=J(t)G_\text{loc}(t,t')J(t')$, where  $G_{loc}(t,t')=-i\langle T_{\cal C}c(t) c^{\dagger}(t') \rangle$.

Using slave-rotor decomposition, we represent an impurity electron by $c_{\sigma}^{\dagger}=f_{\sigma}^{\dagger}e^{i\theta}$, where
$\theta$ is a phase defined in $[0,2\pi)$, associated with an angular momentum $L=-i \partial/\partial \theta$ and $f_{\sigma}$, is a fermionic charge-less operator with spin $\sigma$, known as spinon. In this representation, the 
the projection to the physical Hilbert space
is imposed by
\begin{equation}
L = \sum\limits_{\sigma} \Big(f^{\dagger}_{\sigma} f_{\sigma} -\frac{1}{2} \Big),\label{Eq:constcharge},
\end{equation}
and the commutator relations
\begin{align}
 &[ L(t), \theta(t') ] = -i \delta_{\cal C}(t,t'),\label{Eq:stattheta} \\
 &\{ f_{\sigma}(t), f_{\sigma'}^{\dagger}(t') \} = \delta_{\sigma \sigma'}\delta_{\cal C}(t,t'),\label{Eq:statf} 
\end{align}
assure the electron statistics.
It is clear from Eqs.~(\ref{Eq:constcharge}, \ref{Eq:stattheta}), that $\theta$ is a bosonic field which carries the charge information.

Using slave-rotor representation, we express the impurity action of Eq.~\eqref{Eq:actionelectron} as
\begin{align}
 \mathcal{S}=&  -i\Big\{  \int_{\cal C} {\rm dt dt'} f_{\sigma}(t)e^{-i\theta(t)} \Delta(t,t') f_{\sigma}^{\dagger}(t')e^{i\theta(t')}, \nonumber \\ 
	    & + \int_{\cal C} {\rm dt} U  L^{2}
	    + \int_{\cal C} {\rm dt} L(t) (-\partial_{t}) \theta(t)   \nonumber \\
	    & + \int_{\cal C} {\rm dt} \sum\limits_{\sigma} f_{\sigma}^{\dagger}(t) (-i\partial_{t}) f_{\sigma} (t) \nonumber \\
	   & - \int_{\cal C} {\rm dt} \lambda \Big( L(t)- \sum\limits_{\sigma}f_{\sigma}^{\dagger}(t) f_{\sigma}(t)  -1 \Big) \Big\}
	    ,\label{Eq:actionthetaspinon}
\end{align}
where $\lambda$ is the time-independent Lagrange-multiplier. To achieve a $L$-free action, we first replace operators by their corresponding fields and then integrate over $L$ fields.

After introducing a quantum rotor as $X=e^{i\theta}$, 
with the constrains as $|X|^{2}=1$,  which is imposed by a time-dependent Lagrange-multiplier $\eta(t)$,  we obtain (ignoring constant terms)
\begin{align}
 \mathcal{S}=&  -i\Big\{  \int_{\cal C} {\rm dt dt'} f_{\sigma}(t)X^{*}(t) \Delta(t,t') f_{\sigma}^{*}(t')X(t'), \nonumber \\ 
	    & + \int_{\cal C} {\rm dt}  \big( -i\partial_{t} +\lambda \big) X(t) \frac{1}{4U}  \big( i\partial_{t} +\lambda \big) X^{*}(t) \nonumber \\
	    & + \int_{\cal C} {\rm dt} \sum\limits_{\sigma} f_{\sigma}^{*}(t) (-i\partial_{t} - \lambda) f_{\sigma} (t) \nonumber \\
	   &-\eta \int_{\cal C} {\rm dt} X(t) X^{*}(t)
	   \Big\} 
	    ,\label{Eq:actionslaverotor}
\end{align}

The impurity model can be solved exactly when the spin-degeneracy $N$ and the number of rotor flavors $M$ is increased from $N=2$ and $M=1$ to infinity, keeping the ratio $\mathcal{N}=N/M$ fixed \cite{Florens2002},  and this limit provides a qualitatively correct description of the metal-insulator transition. (In our presented results, we fix the parameter $\mathcal{N}= 3$, for which the DMFT phase-diagram is quantitatively reproduced \cite{Florens2002}.) 
At half-filling, we set $\lambda$ to zero, and acquire the Dyson equations of spinon and rotor fields, using Eq.~\eqref{Eq:actionslaverotor}, for 
\begin{align}
\label{GX}
 G_{X}(t,t')&=-i\langle T_{\cal C}X(t) X^{*}(t') \rangle ,\\
 \label{Gf}
 G_{f}(t,t')&=-i\langle T_{\cal C}f_{\sigma}(t)f^{*}_{\sigma}(t') \rangle,
\end{align}
as
\begin{align}
\label{dysonf}
 &\big(
i\partial_{t}-\mu 
\big) G_{f}(t,t')
-
[
\Sigma_{f}*G_{f}
](t,t')
=
\delta_{\cal C}(t,t'),\\
&\Big(
\frac{-1}{4U}\partial_{t}^{2}
+\eta
\Big)
G_{X}(t,t')
-
[
\Sigma_{X}*G_{X}
](t,t')
=
\delta_{\cal C}(t,t'),
\label{dysonx}
\end{align}
where $ \Sigma_{X}(t,t')=i{\cal N}\Delta(t,t') G_{f}(t',t)$ and $\Sigma_{f}(t,t')=i\Delta(t,t') G_{X}(t',t)$, are rotor and spinon self-energies.

After solving Dyson equations and computing the rotor and spinon's Green's functions, the electron's Green's function is obtained by $G_{loc}(t,t')=iG_{f}(t,t')G_{X}(t,t')$, closing the equations with the DMFT self-consistency. Equations \eqref{dysonf} and \eqref{dysonx} are solved using the Volterra integral techniques described in Ref.~\cite{Aoki2014}, and $\eta(t)$ is determined  by a predictor-corrector procedure.

\section{Anderson Model without lattice self-consistency}

In this section we use the nonequilibrium slave rotor impurity solver to study a fast quench in the single-impurity Anderson model, i.e. the impurity problem of the Hubbard model without DMFT self-consistency. The results will confirm that the slow quasiparticle formation in the Hubbard model, which is described in the main manuscript, is indeed a lattice effect and not a property of the DMFT impurity model alone. We choose an Anderson impurity model at $U=2$, with hybridization function $\Delta(t,t')=J(t)G_0(t,t')J(t')$, where $G_0$ is the local Green's function of a bath with semi-elliptic density of states of bandwidth $4$ and  inverse temperature $\beta=65$, and the coupling is ramped from 
$J(t)=0$ for $t<0$ to $J(t)=0.4$ for $t>0.25$. In equilibrium (time-independent $J=0.4$), this parameter regime corresponds to a three-peak structure of the spectral function, where the central (Kondo) peak has a width of approximately $0.25$.

\begin{figure}[tbp]
\centerline{ \includegraphics[width=0.99\columnwidth,clip=true]{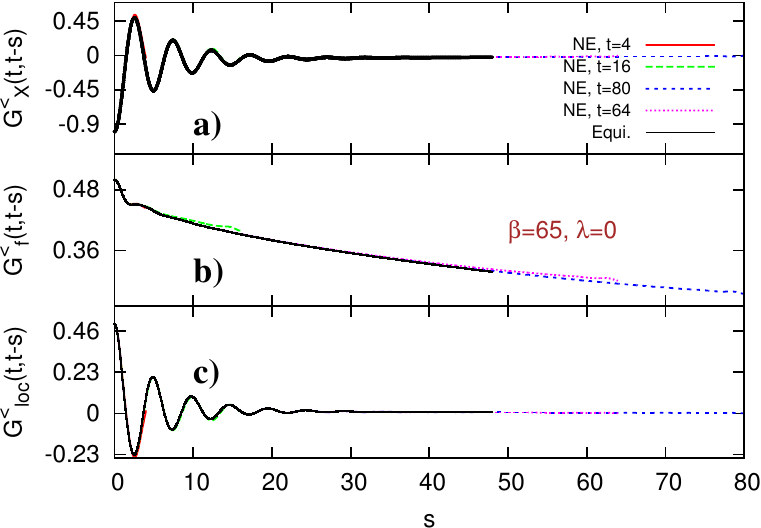}}
 \caption{
 Lesser components of rotor(a), spinon(b), and electron(c) at three different times $t \in \{16,64,80\}$, as well in equilibrium, as a function of time differences. (Anderson impurity model with $U=2$ and $\beta=65$, see main text.)
 }
\label{fig:IMP_GF_Gloc_GX}
\end{figure}

In Fig.~\ref{fig:IMP_GF_Gloc_GX}, we plot the electron, spinon and, rotor lesser Green's functions in and out of equilibrium. In contrast to the corresponding behavior in the Hubbard model (c.f. Fig. 3a of the main text), all Green's functions are time-translational invariant for 
short 
times $t\gtrsim 16$. In addition, the equilibrium results (black points) of the electronic Green's function lie on top of our nonequilibrium data, indicating that the system is 
equilibrated. This confirms that the spectral functions corresponding to the Kondo peak are retrieved without any apparent bottleneck behavior. 

These results also provide a test for the slave-rotor impurity solver, as they are consistent with previous investigations of the dynamics in the Anderson model \cite{Nordlander1999}, which show that the buildup of the Kondo-peak is limited by energy-time uncertainty, i.e., the formation of the peak is complete after a timescale $t_K\sim 1/T_K$, where $T_K$ is the width of the peak.

\section{Solution of the problem using the noncrossing approximation}

\begin{figure}[tbp]
\centerline{ \includegraphics[width=0.9\columnwidth,clip=true]{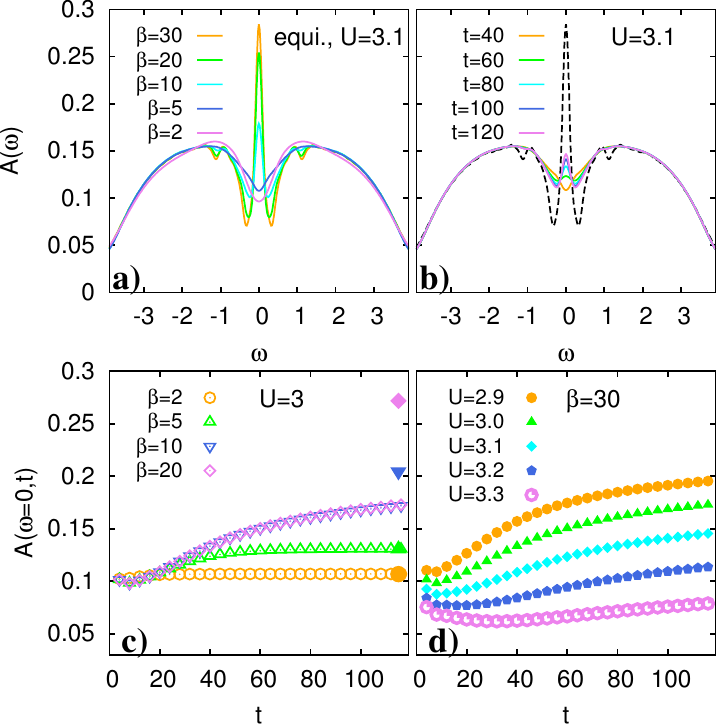}}
 \caption{
Similar as Figure 1 of the main text, but using NCA to solve the DMFT equations:
a) Spectral function in equilibrium for various temperatures $U=3.1$ below the metal insulator transition.
b) Time-dependent spectral function $A(t,\omega)$ after the ramp-on of the hopping, for $U=3.1$ and inverse bath temperature $\beta=30$. The dashed line is the equilibrium result.
c) Spectral weight $A(t,\omega=0)$ at $\omega=0$ as a function of time for $U=3$ and inverse bath temperatures $\beta$ as indicated.
The filled symbols correspond the the height $A(\omega=0)$ of the quasiparticle peak in equilibrium at the same $\beta$.
d) Time dependent spectral weight $A(t,\omega=0)$ at low bath temperature for various interactions $U$.
 }
\label{fig:supp-1}
\end{figure}

\begin{figure}[tbp]
 \centerline{ \includegraphics[width=0.9\columnwidth,clip=true]{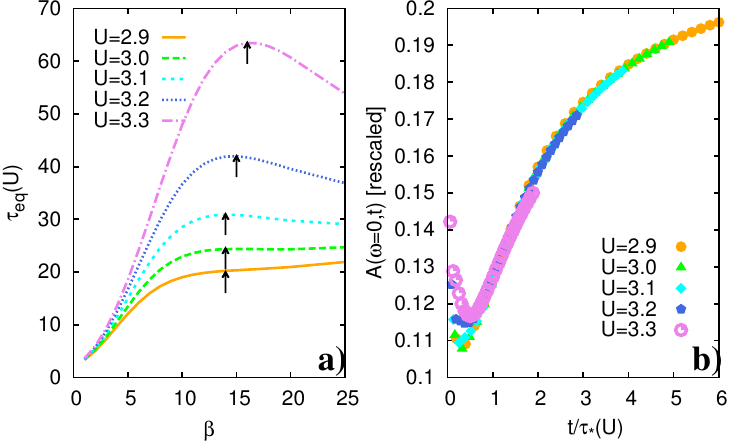}}
 \caption{ a) The crossover scale $\tau_{eq}(U,\beta)$ as a function of inverse temperature, obtained using Eq.~\eqref{a2cosh-supp} with the spectral functions obtained within NCA. The arrows point at the maximal or saturated value, used to define the crossover scale $\tau_*(U)$. The resulting values are $\tau_*(U)=$
$20$ ($U=2.9$), 
$24.5$ ($U=3$), 
$31$ ($U=3.1$), 
$42$ ($U=3.2$), 
$63.5$ ($U=3.3$).
b) 
Time dependent spectral weight $A(\omega=0,t)$ at bath temperature $\beta=30$ for various interactions $U$
(same data as in Fig~\ref{fig:supp-1}d) plotted as a function of rescaled time $t/\tau_*(U)$, with the timescale $t/\tau_*(U)$ taken from panel a). Additionally, the curves are rescaled with an arbitrary factor in vertical axis.
 }
\label{fig:supp-2}
\end{figure}

In this section we analyze the quasiparticle formation for the same setup as in the main text, but instead of the slave-rotor decoupling we now use the lowest order strong-coupling expansion \cite{Eckstein2010} (non-crossing approximation, NCA) to solve the DMFT equations.  (The implementation of the NCA is described in Ref.~\cite{Eckstein2010}.) 
The NCA is known to underestimate the critical interaction $U_c$ of the Mott transition, 
but nevertheless the method qualitatively reproduces the phase diagram of the Mott transition. It is therefore illustrating to see that also the dynamics of quasiparticle formation described in the main text is captured by the NCA solution. 

We use the same setup as in the main text, i.e., the hopping $J(t)$ in the Hubbard model [Eq.~(1) of the main text] is suddenly ramped from the atomic limit $J=0$ to unity $J=1$, and then the equilibration in a thermal bath of temperature $1/\beta$ is studied.   The coupling to the bath is set to $\lambda=0.5$ throughout this section. Figure \ref{fig:supp-1}a) shows the spectral function in equilibrium ($J=1$) for various temperatures at $U=3.1$, which is close to the NCA value for the critical interaction of  the metal-insulator transition. The curves show the crossover from the bad metal at high temperatures to the metal with a quasiparticle peak at low temperatures. 

Figure~\ref{fig:supp-1}b) shows the time-evolution of the spectral function $A(t,\omega) = - \frac{1}{\pi} \text{Im} \int_0^t ds G^{ret}(t-s,t) e^{i\omega s}$ after the ramp-on of the hopping at $t=0$, which reveals a slow recovery of the quasiparticle peak. In Fig.~\ref{fig:supp-1}c) we plot the spectral weight at $\omega=0$, i.e., the height of the quasiparticle peak, for various temperatures $1/\beta$ of the bath. The final equilibrium height of the quasiparticle peak is recovered within the simulation time only if the temperature is above some crossover scale ($\beta=2,5$ in Fig.~\ref{fig:supp-1}c). At lower temperatures the evolution becomes basically independent of the bath temperature $1/\beta$, whereas the quasiparticle peak in equilibrium would strongly increases with decreasing temperature (the data for $\beta=10$ and $\beta=20$ in Fig.~\ref{fig:supp-1}c) fall almost on top of each other). Figure \ref{fig:supp-1}d) shows that the evolution becomes slower as $U$ is increased. All this behavior, which indicates the existence of an electronic bottleneck time related to the metal-insulator crossover regime, is therefore perfectly in agreement with the slave-rotor solution (c.f.~Fig.~1 of the main text), although shifted to smaller values of the interaction due to the underestimation of the critical $U$ within NCA. (The second order strong-coupling approximation would almost quantitatively yield the phase diagram in equilibrium, but the higher numerical effort does not allow to reach the long simulation times needed to systematically analyze the quasiparticle formation.)

Furthermore, we can investigate whether the slowdown of the dynamics in the NCA solution is determined by the electronic bottleneck time $\tau_*$, which has been identified  from the spinon lifetime using the slave-rotor language. Here we extract $\tau_*$ from the electronic spectral functions, using the approximate form  given by Eq. (4) of the main text.  In Fig.~\ref{fig:supp-2}a) we show the timescale $\tau_{eq}$ obtained by using Eq.~(4) of the main text
\begin{equation}
\tau^{-1}_{\rm eq}(\beta)
=
\int d\omega
\,\frac{\pi A(\omega)^2}{\cosh(\beta\omega/2)^2},
\label{a2cosh-supp}
\end{equation}
as a function of temperature $1/\beta$, using the DMFT self-consistency for the semi-elliptical density of states, $\Delta(\omega)''=-\pi A(\omega)$. The maximum or saturation value (black arrows) defines the crossover scale $\tau_*(U)$. In Figure \ref{fig:supp-2}b) we plot the data of Fig.~\ref{fig:supp-1}d) as a function of rescaled time $1/\tau_*(U)$, with an additional rescaling of the vertical axis. We observe an approximate data collapse, which confirms that the time-evolution of the quasiparticle height $A(\omega=0,t)$ for various values of $U$ roughly satisfies a functional form $A(\omega=0,t) = a_Uf(t/\tau_*(U))$ with a given $U$-independent functional form, 
analogous to the Slave Rotor results represented in Fig.~1e of the main text.